\documentstyle[twocolumn,aps,prb]{revtex}

\begin{document}
\draft
\title{Magnetoconductance Oscillations  \\in 
Ballistic Semiconductor-Superconductor 
Junctions
 }
\author{Yasuhiro Asano\footnote{e-mail: asano@eng.hokudai.ac.jp}
}
\address{
Department of Applied Physics, Hokkaido University, Sapporo 060-8628, Japan.\\}
\date{\today}
\maketitle
\begin{abstract}
The mechanism of the magnetoconductance oscillations in junctions of a 
ballistic semiconductor and a superconductor is discussed.
The oscillations appear when both the normal and the Andreev 
reflection occur at the interface. 
The interplay between the classical cyclotron motion 
of a quasiparticle and the phase shift caused by the magnetic field
is the origin of the conductance oscillations.
\end{abstract}

\hspace{0.5cm}


Conductance oscillations as a function of the applied magnetic field in a small ring 
are a one of fundamental consequence of the phase-coherent transport~\cite{webb}. 
The width of the ring must be narrow so that the number of the 
propagation path of an electron wave can be limited.
If many propagation paths are allowed in the ring, 
the magnetoconductance oscillations (MCO) are washed out.
In order to observe the MCO clearly, at least, we have to
either confine the electron wave as in the 
experiment~\cite{webb} or the magnetic flux as in the original idea~\cite{aharonov}.
An electron wave, however, 
can be confined within a classical trajectory of the cyclotron motion 
by its charge degree of freedom under 
the relatively strong magnetic fields, which is called the magnetic 
focusing~\cite{benistant,vanhouten}. 
In the ballistic transport regime, we show a possibility to observe the  
MCO in simply-connected system.
 
In this paper, we discuss the conductance
in small semiconductor-superconductor (Sm-S) junctions, where 
a high mobility two-dimensional electron gas (2DEG) is used 
as a semiconductor.
Recently S-Sm-S junctions can be realized in strong magnetic fields 
~\cite{takayanagi}.
It has numerically shown that the sinusoidal MCO appears in relatively 
weak magnetic fields when the Andreev reflection~\cite{andreev} is not
 perfect~\cite{takagaki}. 
To date, however, the mechanism of the sinusoidal MCO has remained
unclear. 
In this paper, we reveal the nature of the novel MCO within a simple analysis. 
We conclude that the interplay between the cyclotron motion of a
quasiparticle (QP) and the phase shift caused by the magnetic field 
is responsible for the MCO in simply-connected Sm-S junctions. 
The MCO is one of the interference effect of the Andreev
reflected QP, which have been focused 
recently~\cite{vanwees,beenakker,nazarov,morpurgo}.

Let us consider a two-dimensional wire where electrons are confined
in the $y$ direction in the range of $-W/2 < y< W/2$.   
The wire consists of 2DEG ($x<0$) and S ($x>0$).
The Sm-S junctions are described by the 
Bogoliubov-de Gennes equation~\cite{degennes}
\begin{equation}
\left(
\begin{array}{cc}
H_0 & \Delta(x,y) \\
\Delta(x,y)^\ast & -H_0^\ast
\end{array}
\right)
\left(
\begin{array}{c}
u \\
v
\end{array}
\right)
=
E\left(
\begin{array}{c}
u \\
v
\end{array}
\right),\label{bdg}
\end{equation}
where $u(x,y)$ and $v(x,y)$ are the wavefunctions of a quasiparticle.
The Hamiltonian is given by
$ H_0= -{\hbar^2}
\{ \nabla - ie {\bf A}(x,y)/\hbar c\}^2/ {2m^\ast}
+ U(x,y) - \mu $, 
where the mass of an electron $m^\ast$ is $m_N $ for $x \leq 0$ and 
$m_S $ for $x > 0$, respectively. The chemical potential
of the junction is denoted 
by $\mu$. In what follows we set the chemical potential as an 
origin of the energy, i.e., $\mu =0$. 
The Fermi energy in 2DEG and S are $\mu_N$ and $\mu_S$, respectively, 
which correspond to the energy difference between
the band edge and the chemical potential.
The scalar potential $U(x,y)$ involves the hard wall confinement potential
in the $y$ direction and the potential barrier at the Sm-S interface
which is described by $H\delta(x)$. 
The potential barrier height should be determined consistently by the 
the electronic structure on either sides of the junction~\cite{schussler}.
In this paper, however, we treat $H$ as one of the independent parameters
of Sm-S junctions.
We assume that pair potential $\Delta(x,y)$ is $\Delta_0$ 
in S and zero in 2DEG, respectively.
This model is justified when 
the superconducting segment is covered with materials 
with high magnetic permeability, because 
the magnetic field is not applied onto S~\cite{screening}.
Since S is magnetically shielded, 
the vector potential is ${\bf A}=(0,0)$ 
for $x \geq 0$ and ${\bf A}=(0,Bx)$ for $x<0$.
In what follows, we measure the energy and the length in units of $\mu_N$ and 
$1/k_F\equiv \sqrt{2m\mu_N} /\hbar$, respectively.

The wavefunction in 2DEG and that in S
can be obtained separately, and are related with each other by using the continuity 
condition at $x=0$.
The detail of the numerical simulations is given  
elsewhere~\cite{takagaki,tamura}.
Here we show the expression of the zero-bias conductance at zero 
temperature~\cite{takane},
$G = ({2e^2}/{h}){\sum_{l,n}}' 
\left( \delta_{l,n}- R_{l,n}^{ee} +R_{l,n}^{he} \right)$, 
where $l$ and $n$ label the propagation channels in 2DEG under the magnetic field, 
$R_{l,n}^{ee}$ and $R_{l,n}^{he}$ 
 are the reflection probability 
into the electron and hole branches with $E\rightarrow 0$, respectively.
The summation ${\sum_l}'=N_c$ runs over the all propagating channels. 
In the limit of $E\rightarrow 0$, there is no propagating channel in S.
The current conservation low implies
${\sum_l}'( R_{l,n}^{ee} +R_{l,n}^{he} ) = 1$.

In Fig.~\ref{bdep}, we show the numerical results of the
conductance in units of $2e^2/h$ as a function of
$\beta\equiv\mu_N/\hbar\omega_c$, where $\omega_c=eB/c m_N$ 
and $\Delta_0/\mu_N=0.02$, respectively.
Since the pair potential in S is typically 1 meV and 
the Fermi energy in 2DEG is about 100 meV, we fix 
$\Delta_0/\mu_N$ at 0.02 throughout this paper. 
We also fix the width of the wire $W k_F$ at 40.
The numerical results of the conductance presented here are essentially the same with 
those in Ref.~\onlinecite{takagaki}.
There are three parameters which characterize the Sm-S junctions 
and the reflection probability at the interface:(i) the difference 
of the Fermi energy, $\mu_S/\mu_N$, (ii) the difference 
of the effective mass, $m_S/m_N$, and (iii) the potential barrier, 
$V_{bh}\equiv 2m_NH/\hbar^2k_F$. 
In the the solid line of Fig.~\ref{bdep}(a), 
we show the conductance for $\mu_S/\mu_N=1$, $m_S/m_N=1$ and 
$V_{bh}=0$.
The results show the conductance step and the conductance decreases with 
increasing the magnetic field.
In this case, we have confirmed that the normal reflection does not occur 
at the interface from numerical data, i.e., $R^{ee}_{l,n}\simeq 0$.
This can be also understood by the S-matrix in one-dimensional Sm-S 
interface 
\begin{eqnarray}
r_{ee} &=& {\sqrt{(1-|\xi|^2)^2+(2{\rm Im}\xi)^2}}/({1+|\xi|^2}){\rm e}^{i\theta_n} 
= r_{hh}^\ast , \label{1dree}\\
r_{he} &=& 2{{\rm Re}\xi}/({1+|\xi|^2}){\rm e}^{-i\pi/2} = r_{eh}, 
\label{1drhe}
\end{eqnarray}
where $\xi \simeq \sqrt{ ({m_N}/{m_S})({\mu_S}/{\mu_N})} +iV_{bh}$
and $\tan\theta_n \simeq  { 2 V_{bh} }/[
( {m_N}/{m_S}) \cdot ({\mu_S}/{\mu_N})-1 
+ V_{bh}^2 ]$.
Here we solve Eq.~(\ref{bdg}) in one-dimension 
and calculate the reflection
coefficients in the limit of
$\Delta_0 \ll \mu_N$ and $E=0$.
In the present situation, $\xi=1$ leads to $|r_{ee}|=0$, which 
is equivalent to previous results~\cite{blonder}.
By using $R^{ee}_{l,n}=0$ and the current conservation low, 
the conductance results in
$G = (4e^2/h) N_c$, where $N_c$ is plotted with the dot line in (a).
Next we take into account the 
difference of the effective mass in (b) and that of the 
Fermi energy in (c), respectively.
The results show the sinusoidal MCO and the conductance 
is at its minimum when $\beta$ is an integer.
In (b) and (c), the Andreev reflection is no longer perfect, 
which can be also explained by the S-matrix in one-dimensional 
Sm-S junctions, i.e., $\xi\neq 1$.
In one-dimensional junctions, the normal reflection
can be expected to be zero when $\mu_S/\mu_N = m_S / m_N$ and $V_{bh}=0$.
The situation is almost the same even in the 
two-dimensional junction. 
We show the conductance for $\mu_S/\mu_N = m_S / m_N=2.0$ 
in (a) with the dash line.
The MCO disappears and the weak conductance
step again can be seen as well as the solid line.
In Fig.~\ref{bdep}(d), we examine the effects of the potential
barrier at the interface, where $V_{bh}$ = 0.5. 
The potential barrier is one of the origin of the normal reflection 
at the interface. We have confirmed for large $V_{bh}$
that the MCO appears even when $\mu_S/\mu_N$ = $m_S/m_N$.  
Thus we conclude that the MCO appears when both the
normal and the Andreev reflection occur at the Sm-S interface,
(i.e., $||r_{ee}|^2-|r_{he}|^2|^2 \ll 1$), irrespective
of the origin of the normal reflection.
We should note that the S-matrix in the one-dimensional junctions well explains 
the characteristic feature of the normal and Andreev reflection in the 
two-dimensional systems although the dimensionality is different 
in the two systems.
The range of the magnetic fields 
which satisfy the equation 
\begin{equation}
{W}/{2} < L_c < W, \label{condb}
\end{equation}
is denoted by $\leftrightarrow$ in Fig.~\ref{bdep}(b)-(d),
where $L_c \equiv {4}\beta /{k_F}$ 
is the diameter of the cyclotron orbit.
The MCO can be seen   
when Eq.~(\ref{condb}) holds, which 
indicates that the cyclotron motion of a QP 
plays an important role in the MCO.
We have confirmed that the oscillations range shifts to the 
lower magnetic fields as increasing the wire width, 
which has been also reported in Ref.~\onlinecite{takagaki}.   
In fact, we calculate the amplitude of the wavefunction 
reflected into the hole branch, $P_h(x,y)$, as shown in Fig.~\ref{gray1}.
Here we focus on the system in Fig.~\ref{bdep}(d) and 
fix the magnetic field at $\beta$=8.0 in (a) and 5.5 in (b),
respectively.
In the dark area, the wavefunction has the large amplitudes.
An electronlike QP is injected from the lower left corner.
It is important that an incident wave does not have uniform distribution 
at the Sm-S interface but has the large amplitudes at several points, 
as indicated by the arrows.
The classical cyclotron orbits 
are represented by the circles which are drawn to fit the interference 
pattern for $y<0$ and to pass the reflection points at the interface.
It is possible to draw another circles in the figure. We omit them to
avoid complexity. 
Under the magnetic field, it is well known that an electron localizing
near the edge of the wire has the larger velocity in the $x$ direction than
that localizing around the center of the wire. 
Thus at $y/W=-0.4$, the wave is reflected almost perpendicular to the interface,
which corresponds to the fact that the lowest circle is symmetric about $x=0$
as in (a) and (b).
The waves at another reflection points have the velocity in the $y$ direction,
which allows the asymmetric circle about $x=0$.
The figures of Fig.~\ref{gray1} show that the motion of a QP is characterized well 
by the classical cyclotron orbits near the Sm-S interface when Eq.~(\ref{condb}) 
is satisfied.

Based on the numerical results, we make clear the physical picture of the conductance
oscillations within a phenomenological argument.
In the range of the magnetic fields in Eq.~(\ref{condb}),
an incident QP from 2DEG can be scattered twice 
at the Sm-S interface as shown in Fig.~\ref{explain}.
At first an incident electron is either reflected into 
the electron or the hole branches at ${\bf r}_1$. 
After the ballistic motion along the cyclotron orbit (${\bf r}_N$), 
the quasiparticle in each branch is 
reflected again into the electron and hole branches at ${\bf r}_2$. 
Thus an incident QP
divides into four parts as shown in Fig.~\ref{explain}.
In (1) and (3) ( (2) and (4) ), a reflected QP goes across the wire in the electron 
(hole) branch. The phase of a QP is changed by the magnetic field 
while traveling along ${\bf r}_N$. 
When a QP is in the electron branch, the phase change due to the magnetic field 
is given by
$
\phi_B = ({e}/{\hbar c})\int_{{\bf r}_1}^{{\bf r}_2} d{\bf r}_N\cdot
 {\bf A}({\bf r}_N)  = -\pi \beta$.
The phase change of a QP in the hole branch is given by $-\phi_B$.
In the following, we separate the reflection process into three steps.
We describe the two reflections at the Sm-S interface by using 
the S-matrix in one-dimensional junctions.
The effects of the two-dimensionality 
and those of the magnetic fields are taken into account through the 
phase shift by the magnetic field.
In this way, we estimate the wavefunction of the four parts as follows,
\begin{eqnarray}
\Psi_1^e &=& |r_{ee}|{\rm e}^{i\theta_n} 
\cdot {\rm e}^{i\phi_B} \cdot 
|r_{ee}|{\rm e}^{i\theta_n}, \label{p1}\\
\Psi_2^e &=& |r_{eh}| {\rm e}^{-i\pi/2}\cdot 
{\rm e}^{-i\phi_B} \cdot |r_{he}| {\rm e}^{-i\pi/2},  \label{p2}\\
\Psi_3^h &=&  |r_{he}|{\rm e}^{-i\pi/2}
\cdot {\rm e}^{i\phi_B} \cdot |r_{ee}|{\rm e}^{i\theta_n},  \label{p3}\\
\Psi_4^h &=& |r_{hh}| {\rm e}^{-i\theta_n} \cdot {\rm e}^{-i\phi_B} \cdot 
|r_{he}| {\rm e}^{-i\pi/2}.  \label{p4}
\end{eqnarray}   
The two parts in the electron branch 
interfere with each other and $|\Psi_1^e+ \Psi_2^e|^2$ represents the reflection
probability as an electron. In the same way, $|\Psi_3^h+ \Psi_4^h|^2$ is the
reflection probability as a hole.

In Fig.~\ref{explain}, we have assumed that the motion of a QP 
near the interface is characterized by the single cyclotron orbit. 
The amplitude of the reflected wave in Fig.~\ref{gray1}, however, 
show a number of the circles.
When the magnetic field is weak in Fig.~\ref{gray1}(a),
only the trajectory drawn with the solid circle can contribute
to the MCO. 
A QP can return to the
interface after the cyclotron motion along the solid circle. While
another circles go across the wire wall before reaching the interface.
Thus we can neglect the contribution of these orbits to the MCO.
When the magnetic field is relatively strong in Fig.~\ref{gray1}(b),
most of the reflected wave can contribute to the MCO.
The amplitude of the MCO increases with decreasing $\beta$ as shown 
in Fig.~\ref{bdep}.
Within the range of the magnetic fields in Eq.~(\ref{condb}),
the contribution ratio, $p(\beta)$,  
is set to be unity at $W/2 = L_c$ and is zero at $W = L_c$. 
We approximately describe $p(\beta)$ 
by using a linear function of $\beta$
as 
$p({\beta}) = 2(1- {4}\beta/(Wk_F)).$
The conductance  
can be estimated by
\begin{eqnarray}
G &\simeq & g_0 \left[1- p(\beta)^2|\Psi_1^e+\Psi_2^e|^2 + p(\beta)^2|\Psi_3^h+\Psi_4^h|^2\right]\\
 &\simeq &g_0 [ 1 + 4 |r_{ee}|^2|r_{he}|^2 p( \beta)^2 \cos
 ( 2\pi \beta - 2\theta_n) ].
\label{gfinal}
\end{eqnarray}
where $g_0 = (2e^2/h)N_c$ and we use the relation 
$\left( |r_{ee}|^2 - |r_{he}|^2 \right)^2\ll 1$.
In real space, the cyclotron orbits in Fig.~\ref{explain} do not encircle 
the magnetic flux. However, the phase shift by the magnetic field in $\Psi_1^e$
and $\Psi_3^h$ have the opposite sign to that in $\Psi_2^e$ and $\Psi_4^h$.
Thus the magnetic field causes the interference effect.
In Fig.~\ref{explain}, we only consider the symmetric orbits about $x=0$.
The numerical results show a number of the asymmetric circles.
 In Fig.~\ref{gray1}(b), for instance, 
an incident QP is reflected into the two branches at $y/W=-0.28$.
When the QP travels along $O_1$ in the hole branch, the corresponding 
part in the electron branch travels along $O_2$. 
We note that the $O_1$ and $O_2$ are symmetric with each other about $x=0$.
This is because, only the velocity component perpendicular to the 
interface changes sign in the normal reflection, however, all velocity components
change sign in the Andreev reflection~\cite{benistant}.
It can be easily confirmed that the difference in the phase shift between
$O_1$ in the hole and $O_2$ in the electron branches is equivalent
to $2\phi_B$. Thus the asymmetric orbits contribute to MCO as well 
as the symmetric orbits and the phase shift remains constant independent 
of the incident angle of a QP to the Sm-S interface.
A possibility of the conductance oscillations was 
briefly mentioned in Ref.~\onlinecite{takagaki}. 
A part of the argument, however, was not correct.
In the absence of the potential barrier, 
$\theta_n$ becomes zero and the conductance is at its maximum
when $\beta$ is an integer as shown in Eq.~({\ref{gfinal}).  
The numerical results, however, show that the conductance 
takes its minima at these points.
The sign of the oscillating part is a disagreement 
between the simple analysis and the numerical results. 
Since we do not explicitly consider the two-dimensionality, 
the disagreement may stem from the wavefunction in the $y$ direction.
At present, we can not give a satisfactory explanation of the disagreement. 
In Fig.~\ref{bdep}(b)-(d), we compare Eq.~(\ref{gfinal}) (dash line) with 
the numerical results.
The results show an excellent agreement with each other. 
Here we have to confess that the sign of the second term in
Eq.~(\ref{gfinal}) has been changed from $+$ to $-$ in Fig.~\ref{bdep}.

We conclude that the interplay between the classical cyclotron motion
of a quasiparticle and the phase shift by the magnetic field is the origin 
of the magnetoconductance oscillations. 
The conductance oscillations can be seen even in the simply-connected 
Sm-S junctions because
the charge degree of freedom of an electron restricts 
an electron wave into the classical cyclotron orbit under the magnetic field
and the superconductor opens the hole branch in the 2DEG.
Finally we briefly discuss the possibility to observe the MCO in experiments. 
When the width of the wire is $W =5\times 10^{-6}$ m as it is 
in Ref.~\onlinecite{takayanagi}, 
the MCO can be seen around $B\sim 0.08-0.16$ T.
We have assumed the perfect screening of the magnetic field at S.
In experiments, the screened magnetic field is not necessary to be zero.
In this case, the phase fluctuations in the pair potential are caused by 
the magnetic field. However, the phase fluctuations within the length scale $L_c$
is not so large because the coherence length of S is larger than $L_c$ when
S is type I.
In the absence of the magnetic shielding at S, 
we can show that the MCO is washed out or can be seen as the noise-like 
fluctuations when S is type II. 
The details will be given elsewhere.

The author is indebted to N.~Tokuda, H.~Akera, T.~Kato and 
Y.~Takane for useful discussion.

\begin{figure}
\caption{
The conductance is numerically calculated as a function of the inverse
of the magnetic field. 
The number of the propagating channels in 2DEG, $N_c$, 
is plotted by the dot line in (a). 
In (a), the normal reflection does not occur at the Sm-S interface.
While the normal reflection occurs at the interface and
the conductance oscillations appear in (b), (c) and (d).
In (b)-(d), we compare the analytic results (dash line) with the numerical results.
}
\label{bdep}
\end{figure}
\begin{figure}
\caption{
The amplitudes of the wavefunction reflected into the  
hole branch are shown for the system discussed in Fig. 1(d).
We fix the magnetic field at $\mu_N/\hbar\omega_c=8.0$(a) and 5.5 (b),
respectively. The classical cyclotron orbits
are represented by the circles.
}
\label{gray1}
\end{figure}
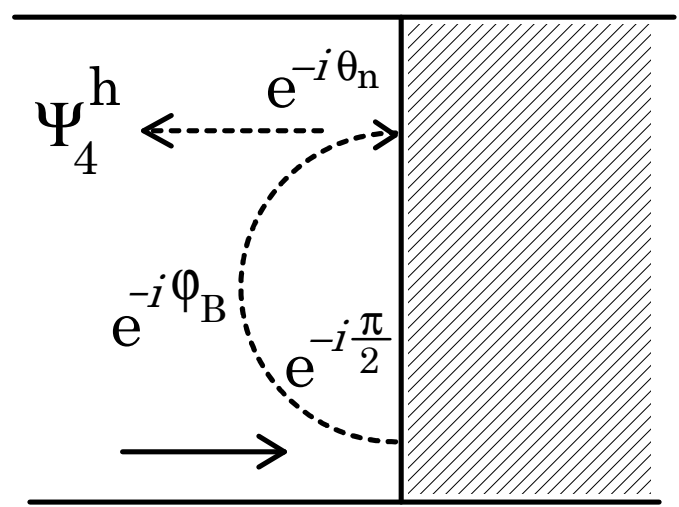
\begin{figure}
\caption{
Schematic picture of the reflection process from the Sm-S interface. 
The solid and broken lines denote
a quasiparticle in the electron and hole branches, respectively.
}
\label{explain}
\end{figure}

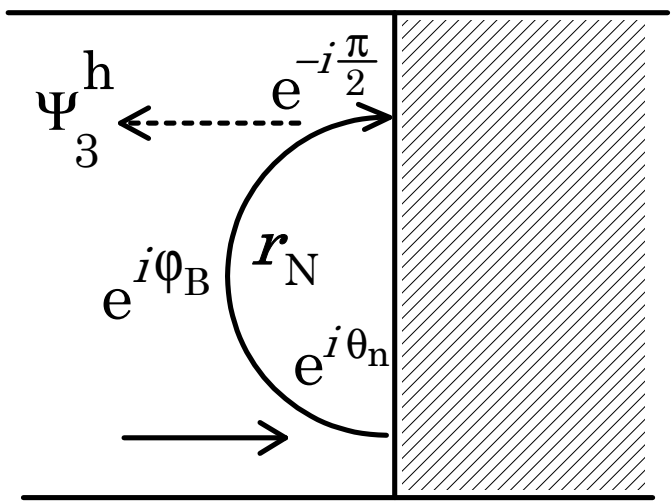
\end{document}